\documentclass[12pt]{iopart}

\usepackage{graphics}
\begin{document}
\title [Comprehensive Analysis of Neutrinos in SK part III]
{
The Comprehensive Analysis of Neutrino Events Occurring inside the
Detector in the Super-Kamiokande Experiment from the View Point of 
the Numerical Computer Experiments: Part~3\\[20pt]
{\large
   -- L/E Analysis for Single Ring Muon Events II --
}
}
\author{
E. Konishi$^1$, Y. Minorikawa$^2$, V.I. Galkin$^3$, M. Ishiwata$^4$,\\
I. Nakamura$^4$, N. Takahashi$^1$, M. Kato$^5$ and A. Misaki$^6$
}

\address{
$^1$ Graduate School of Science and Technology, Hirosaki University, Hirosaki, 036-8561, Japan }    
\address{
$^2$ Department of Science, School of Science and Engineering, Kinki University, Higashi-Osaka, 577-8502, Japan }
\address{
$^3$ Department of Physics, Moscow State University, Moscow, 119992, Russia}
\address{
$^4$ Department of Physics, Saitama University, Saitama, 338-8570, Japan}
\address{
$^5$ Kyowa Interface Science Co.,Ltd., Saitama, 351-0033, Japan }
\address{
$^6$ Research Institute for Science and Engineering, Waseda University, Tokyo, 169-0092, Japan }
\ead{konish@si.hirosaki-u.ac.jp}

\begin{abstract} 
Following $L_{\nu}/E_{\nu}$ analysis in the preceding paper
of the Fully Contained Muon Events resulting from the 
quasi-elastic scattering obtained from our numerical 
computer experiment.
In the present paper, we carry out 
the analyses of $L_{\nu}/E_{\mu}$, $L_{\mu}/E_{\nu}$ and 
$L_{\mu}/E_{\mu}$ among four possible combinations of 
L and E. 
As the result of it, we show that  
we can not find the characteristis of maximum oscillation for 
neutrino oscillation among two of three, 
$L_{\mu}/E_{\mu}$ and $L_{\mu}/E_{\nu}$.
Only  $L_{\nu}/E_{\mu}$ distribution can show
something like maximum oscillation, however it
cannot be detected owing to the neutral 
character of $L_{\nu}$.  
It is, thus, concluded that the Super-Kamiokande Experiment 
could not have found the existence of the maximum oscillation 
for neutrino oscillation.

\end{abstract}
\pacs{ 13.15.+g, 14.60.-z}
\noindent{\it Keywords}: 
Super-Kamiokande Experiment, QEL, Numerical Computer Experiment


%
\maketitle

\section{Introduction}
 In a previous paper \cite{part2}, we have carried out the 
$L_{\nu}/E_{\nu}$ analysis, 
for {\it Fully Contained Muon Events} resulting from the quasi elastic 
scattering(QEL)\cite{Renton} 
obtained from our numerical experiments, namely, the most clear 
cut analysis for the maximum oscillation 
and have shown the existence of the maximum
 oscillations under the neutrino oscillation parameters obtained by 
the Super-Kamiokande Collaboration. 
This fact denotes that our numerical computer experiment has been
 performed in right way.
The maximum oscillations for the neutrino oscillation are derived 
from the survival probability of a given flavor,
 such as, $\nu_{\mu}$,  and it is given by 

\begin{eqnarray}
P(\nu_{\mu} \rightarrow \nu_{\mu}) 
= 1- sin^2 2\theta sin^2
(1.27\Delta m^2 L_{\nu} / E_{\nu} ).  
\end{eqnarray}

\noindent However, as both $L_{\nu}$ and $E_{\nu}$
are not physically measurable quantities which are 
attributed to the nature of neutrino and, consequently,
 the maximum oscillations 
can not be detected through analysis of $L_{\nu}/E_{\nu}$
 distribution, even if they really exist. 
In our numerical computer experiment, we can examine another 
possible combinations of L/E, such as
$L_{\nu}/E_{\mu}$, $L_{\mu}/E_{\nu}$ and 
$L_{\mu}/E_{\mu}$ besides $L_{\nu}/E_{\nu}$.
 Therefore, we try to examine weather the existence of the maximum
 oscillation can be detected through the analysis of L/E besides
$L_{\nu}/E_{\nu}$.

\begin{figure}
\begin{center}
\vspace{-1cm}
\rotatebox{90}{%
\resizebox{0.5\textwidth}{!}{%
  \includegraphics{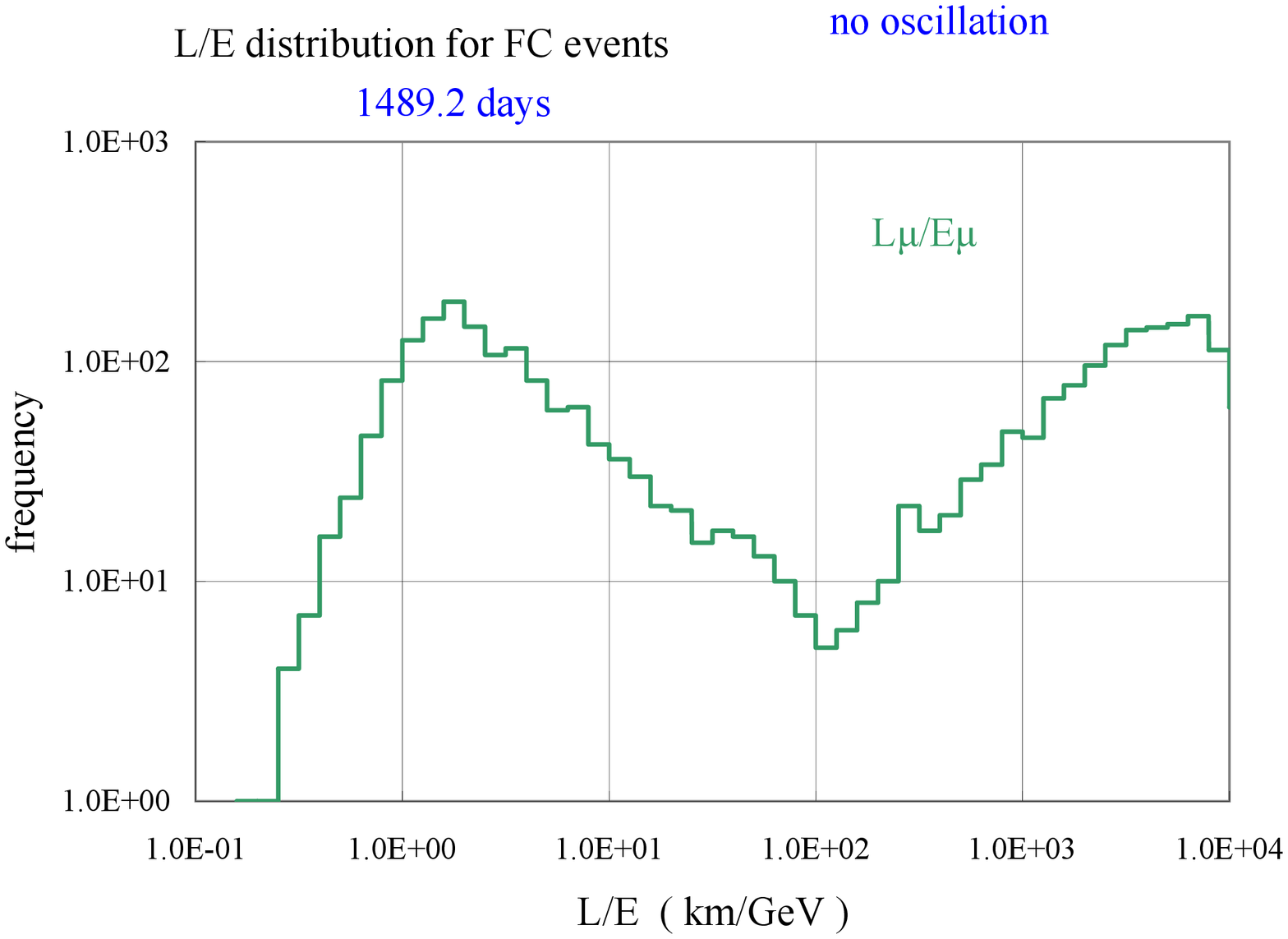}
}}
\vspace{-1cm}
\caption{$L_{\mu}/E_{\mu}$ distribution without oscillation
for 1489.2 live days.}
\label{fig:13}       
\vspace{-1cm}
\rotatebox{90}{%
\resizebox{0.5\textwidth}{!}{%
  \includegraphics{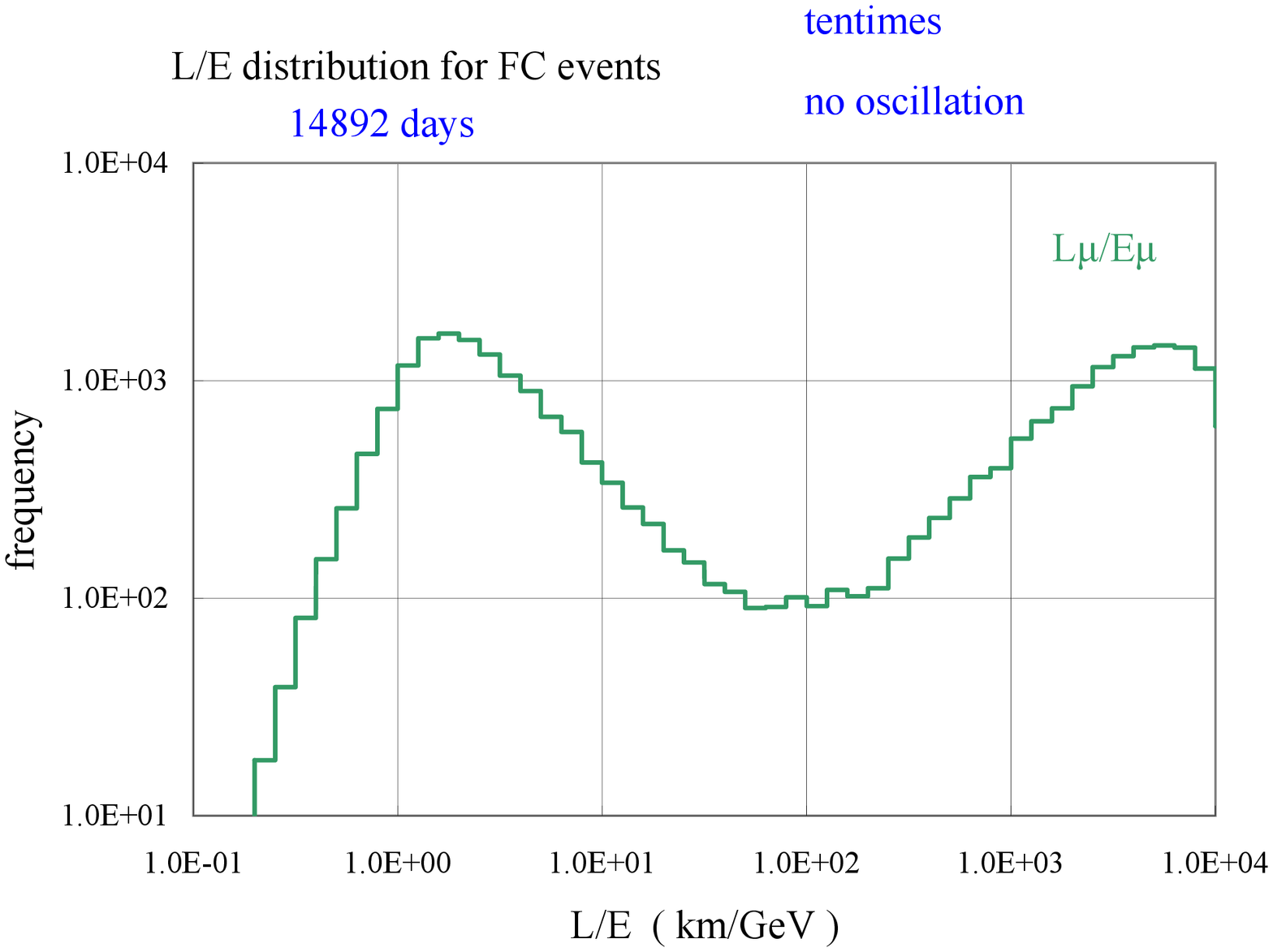}
}}
\vspace{-1cm}
\caption{$L_{\mu}/E_{\mu}$ distribution without oscillation
for 14892 live days.}
\label{fig:14}
\end{center}
\end{figure}

\section{$L_{\mu}/E_{\mu}$, $L_{\mu}/E_{\nu}$
and $L_{\nu}/E_{\mu}$
Distributions in Our Numerical Experiment}

\begin{figure}
\begin{center}
\vspace{-1cm}
\rotatebox{90}{%
\resizebox{0.5\textwidth}{!}{%
  \includegraphics{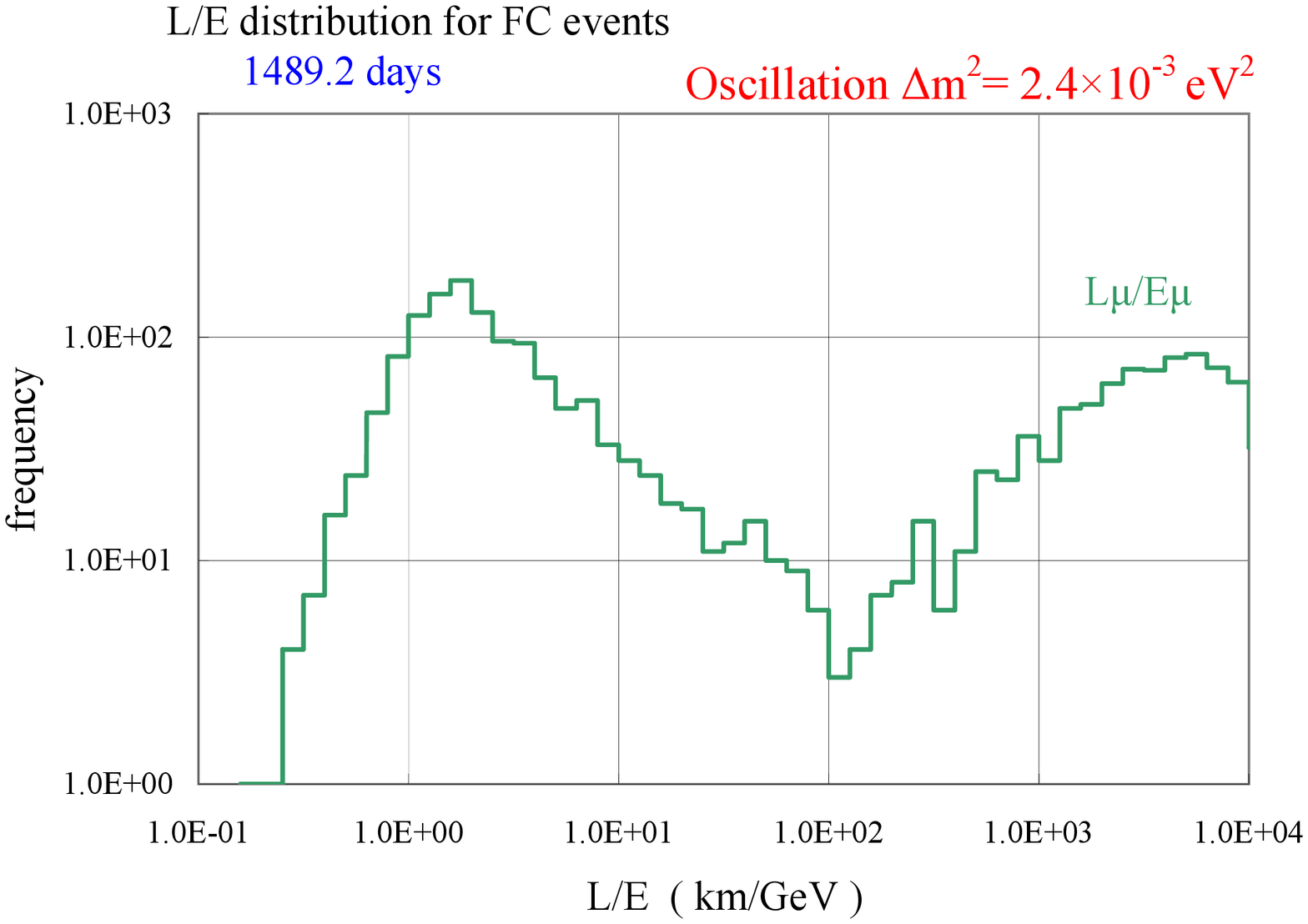}
}}
\vspace{-1cm}
\caption{$L_{\mu}/E_{\mu}$ distribution with the oscillation
for 1489.2 live days.}
\label{fig:15}       
\vspace{-1cm}
\rotatebox{90}{%
\resizebox{0.5\textwidth}{!}{%
  \includegraphics{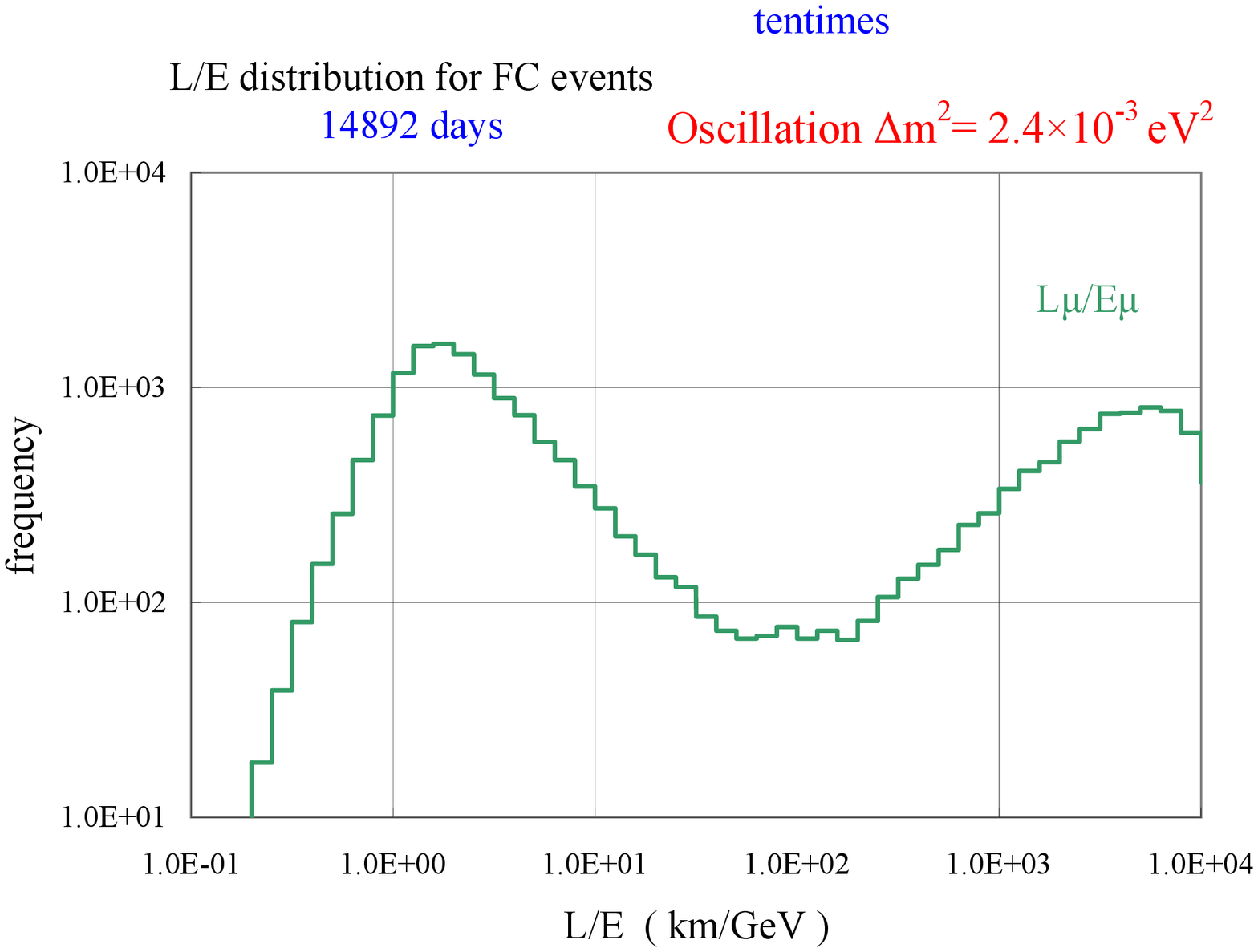}
}}
\vspace{-1cm}
\caption{$L_{\mu}/E_{\mu}$ distribution with the oscillation
for 14892 live days.}
\label{fig:16}
\end{center}
\end{figure}

\subsection{$L_{\mu}/E_{\mu}$ Distribution}
As physical quantities which can really be observed are
$L_{\mu}$ and $E_{\mu}$ instead of $L_{\nu}$ and $E_{\nu}$,
therefore we examine $L_{\mu}/E_{\mu}$ distribution.

\begin{figure}
\begin{center}
\vspace{-1cm}
\rotatebox{90}{%
\resizebox{0.5\textwidth}{!}{%
  \includegraphics{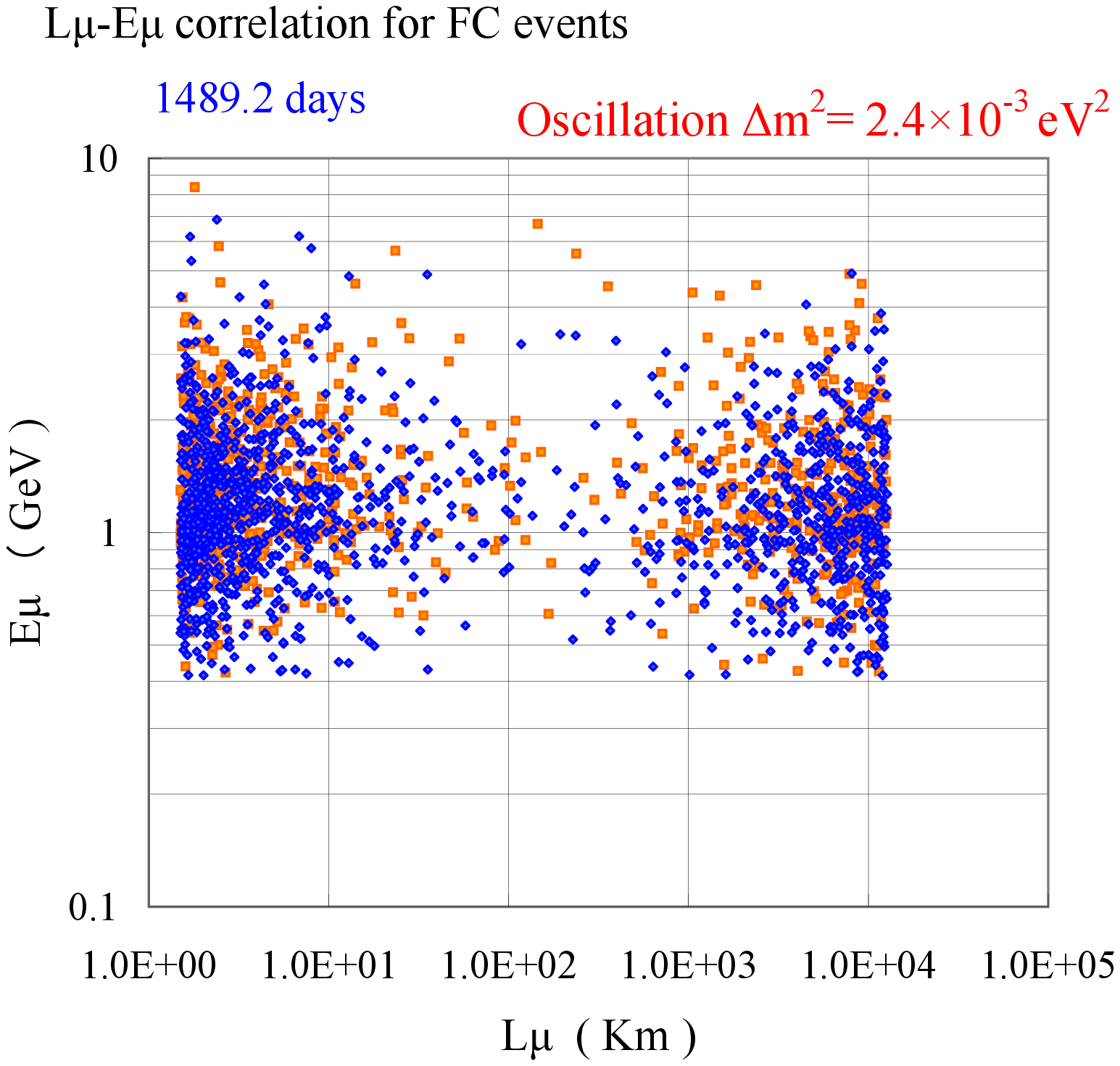}
}}
\vspace{-1cm}
\caption{The correlation diagram between $L_{\mu}$ and $E_{\mu}$
with oscillation for 1489.2 live days.}
\label{fig:17}
\vspace{-1cm}
\rotatebox{90}{%
\resizebox{0.5\textwidth}{!}{%
  \includegraphics{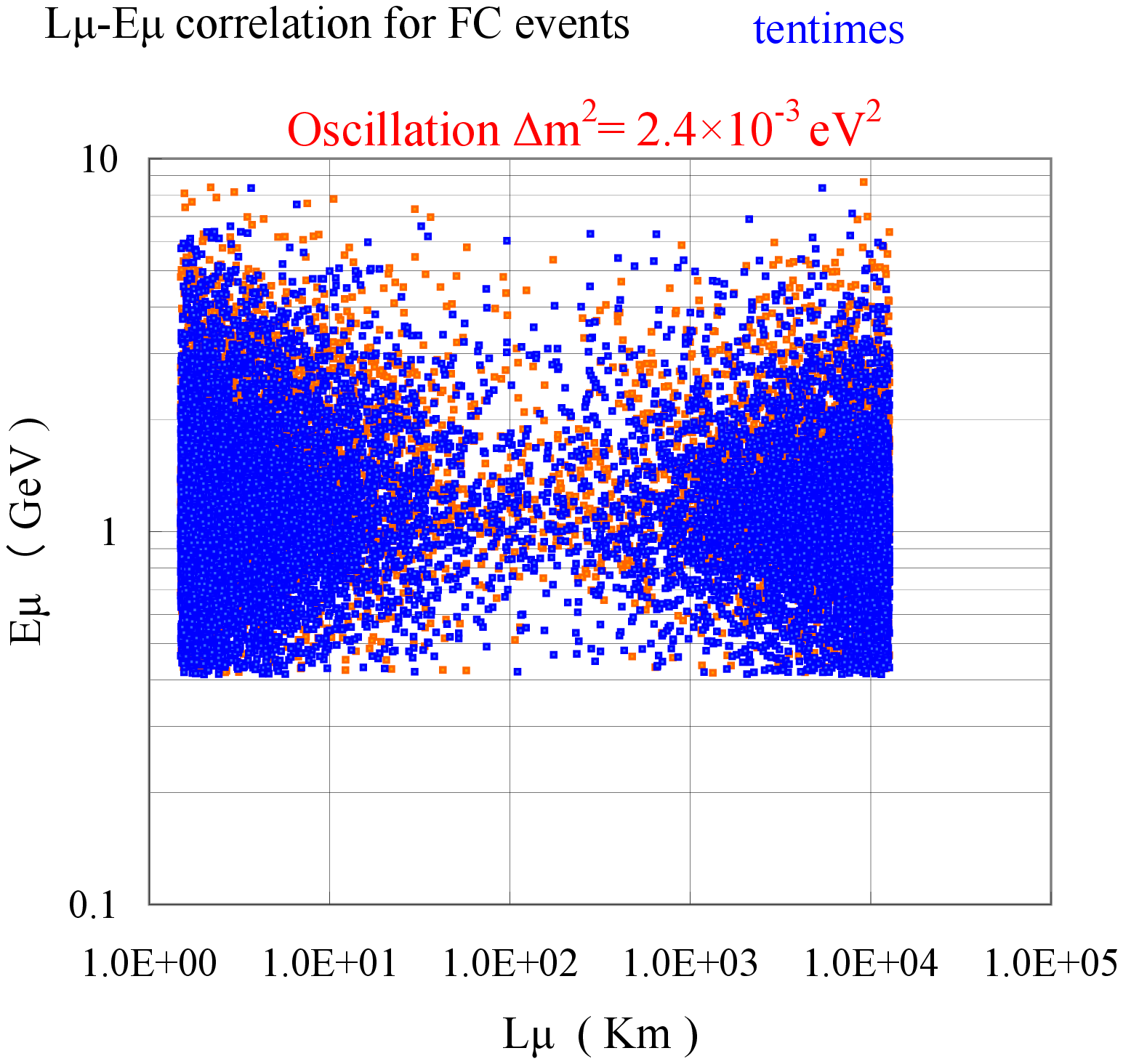}
}}
\vspace{-1cm}
\caption{The correlation diagram between $L_{\mu}$ and $E_{\mu}$
with oscillation for 14892 live days.}
\label{fig:18}
\end{center}
\end{figure}
\begin{figure}
\begin{center}
\vspace{-1cm}
\rotatebox{90}{%
\resizebox{0.5\textwidth}{!}{%
  \includegraphics{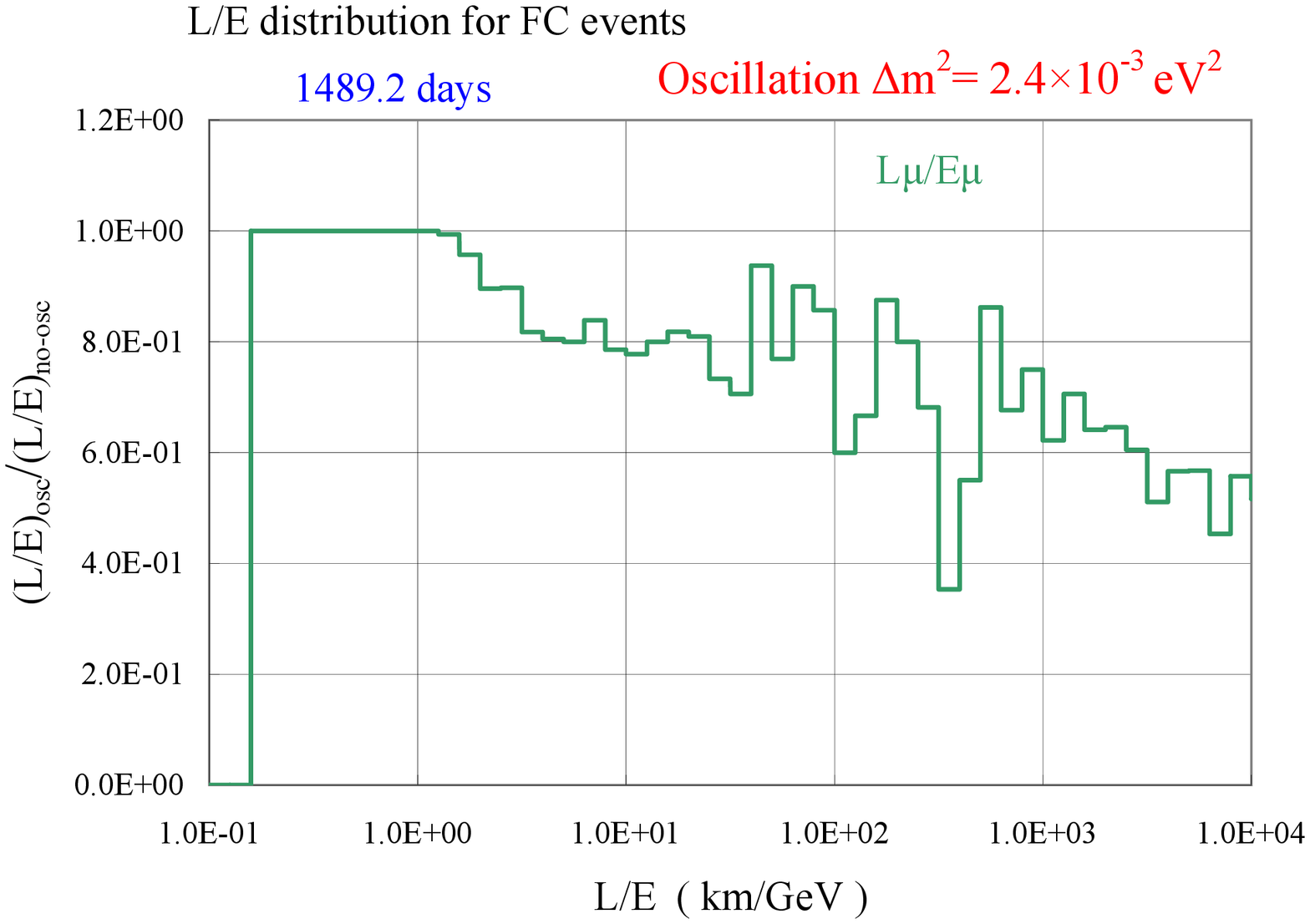}
}}
\vspace{-1cm}
\caption{The ratio of 
$(L_{\mu}/E_{\mu})_{osc}/(L_{\mu}/E_{\mu})_{null}$
 for 1489.2 live days.}
\label{fig:115}       

\vspace{-1cm}
\rotatebox{90}{%
\resizebox{0.5\textwidth}{!}{%
  \includegraphics{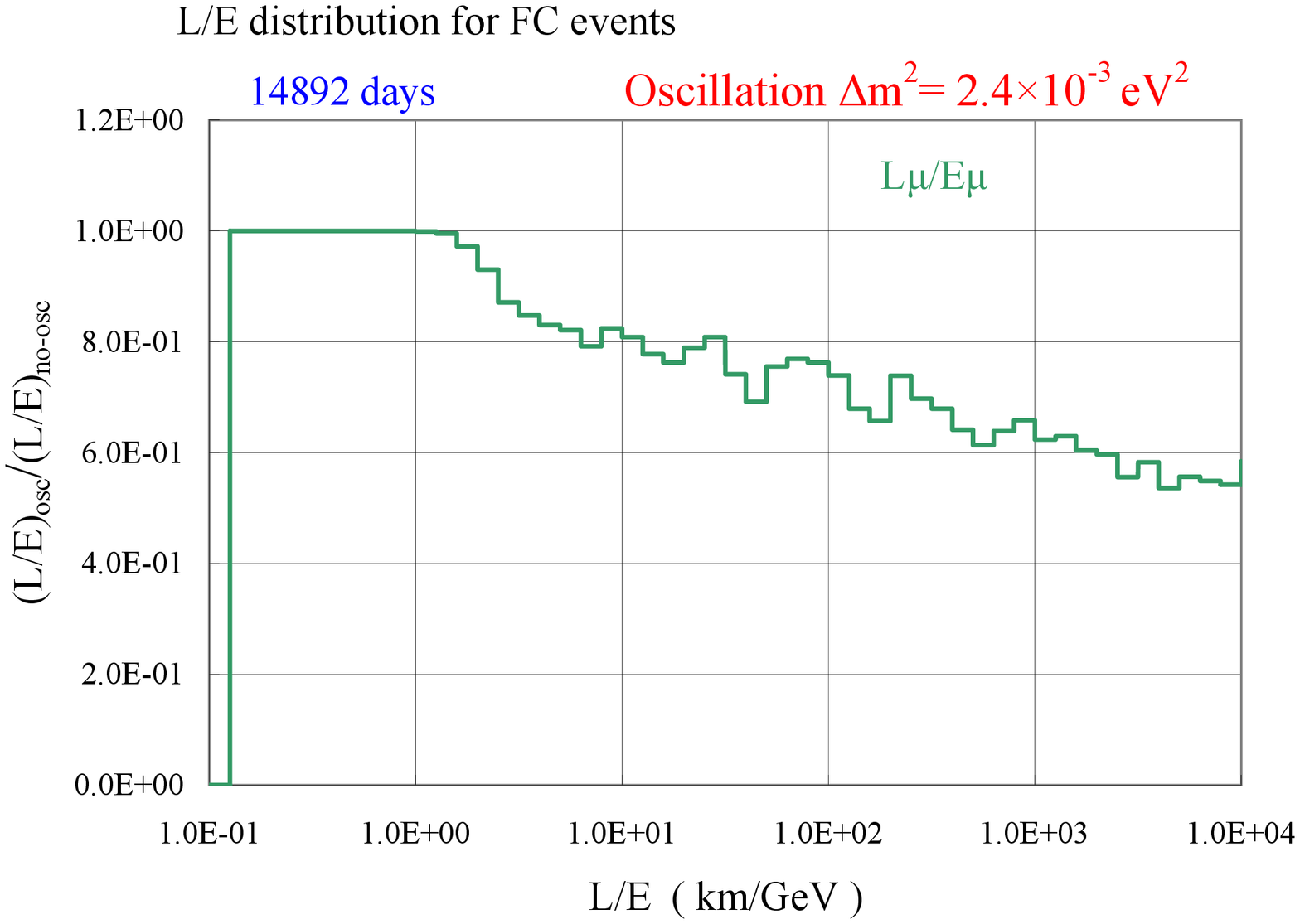}
}}
\vspace{-1cm}
\caption{The ratio of 
$(L_{\mu}/E_{\mu})_{osc}/(L_{\mu}/E_{\mu})_{null}$
 for 14892 live days.}
\label{fig:125}       
\end{center}
\end{figure}

\subsubsection{For null oscillation}
In Figures~1 and 2, we give the $L_{\mu}/E_{\mu}$  distributions 
without oscillation for 1489.2 live days
which is equal to the actual live days
of the Super-Kamiokande Experiment\cite{Ashie2}
 and 14892 live days, ten times as much as that of 
Super-Kamiokande Experiment, respectively.  
Similarly, Figures~1 and 2 show sinusoidal-like 
character as in Figures~6 and 7 for $L_{\nu}/E_{\nu}$
in the preceeding paper\cite{part2}
which has no relation with the oscillation, however.
Such the sinusoidal character represents the intersection 
effect due to the horizontal-like incident neutrino, 
partly the upward neutrinos and partly the downward neutrinos. 
Comparing Figure~1 with Figure~2, the characteristics of the uneven 
histogram in Figure~1 disappear in Figure~2 due to 
ten times statistics as much as that of the Figure~1.
 
\begin{figure}
\begin{center}
\vspace{-1cm}
\rotatebox{90}{%
\resizebox{0.5\textwidth}{!}{%
  \includegraphics{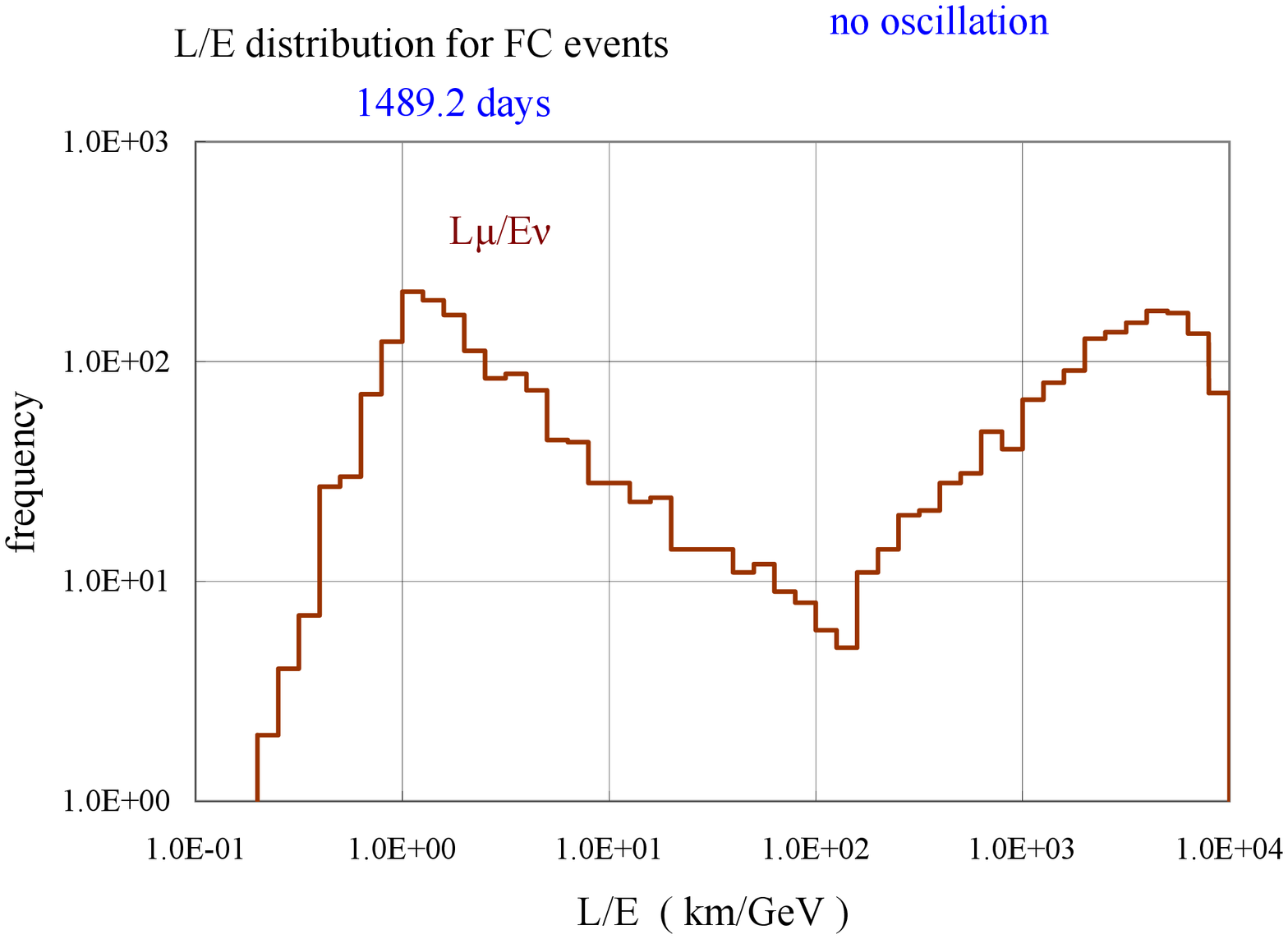}
}}
\vspace{-1cm}
\caption{$L_{\mu}/E_{\nu}$ distribution without oscillation
for 1489.2 live days.}
\label{fig:19}
\vspace{-1cm}
\rotatebox{90}{%
\resizebox{0.5\textwidth}{!}{%
  \includegraphics{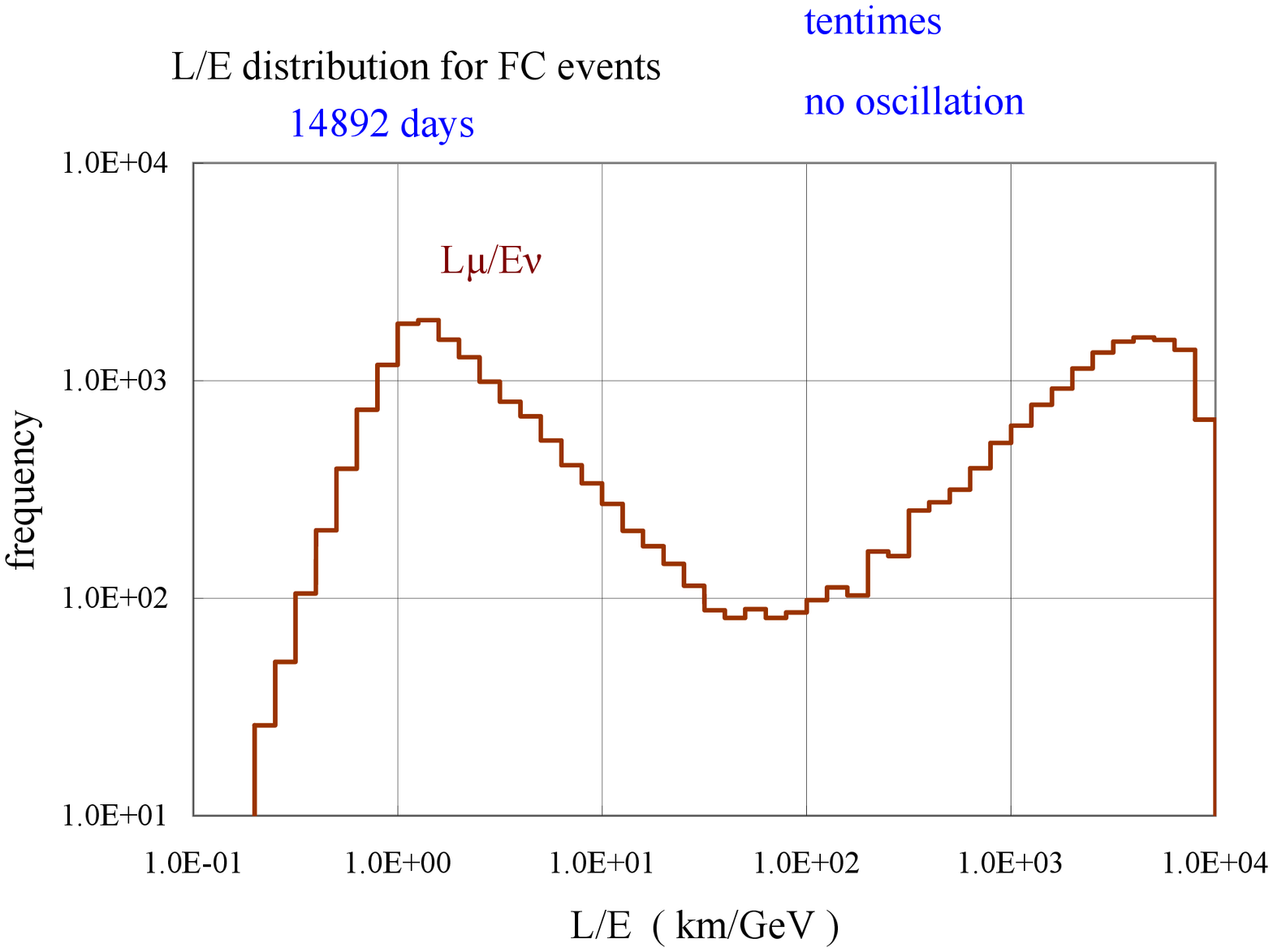}
}}
\vspace{-1cm}
\caption{$L_{\mu}/E_{\nu}$ distribution without oscillation
for 14892 live days.}
\label{fig:20}
\end{center}
\end{figure}
\subsubsection{For the oscillation}
In Figures~3 and 4, we give the $L_{\mu}/E_{\mu}$ distributions with 
the oscillation for 1489.2 live days and 14892 live days, respectively. 
In Figure~3, we may observe the uneven histogram, something like
 dips coming from neutrino oscillation. However, in Figure~4 where 
the statistics is ten times as much as that of Figure~1, the histogram
 becomes smoother and such the dips disappear, which 
turns out finally for the dips to be pseudo. 
Furthermore, comparing Figure~4
 in the presence of neutrino oscillation with Figure~2 in the absence 
of neutrino oscillation, it is clear that the dips which show maximum 
oscillation in the Figure~10
in the preceeding paper\cite{part2}
 are lost in the Figure~4 under cover of the 
complicated relation between $L_{\nu}$ and $L_{\mu}$. 
It is impossible to extract the neutrino oscillation parameters from the 
comparison of Figure~4 with Figure~2.\\ 
In Figures~5 and 6, correspondingly, we give the correlation 
between $L_{\mu}$ and $E_{\mu}$ for 1489.2 live days and 14892 live 
days, respectively.
 It is clear from the figures that we can not
observe any combination of $L_{\mu}/E_{\mu}$ which gives the maximum
 oscillation on the contrary to Figures~11 and 12
in the preceeding paper\cite{part2}.
 Namely, we may conclude that we can not observe the 
sinusoidal flavor transition probability of neutrino oscillation
against the claim by the Super-Kamiokande Collaboration\cite{Ashie1}
when we adopt physically observable quantities, such as
 $L_{\mu}$ and $E_{\mu}$.\\
\begin{figure}
\begin{center}
\vspace{-1cm}
\rotatebox{90}{%
\resizebox{0.5\textwidth}{!}{%
  \includegraphics{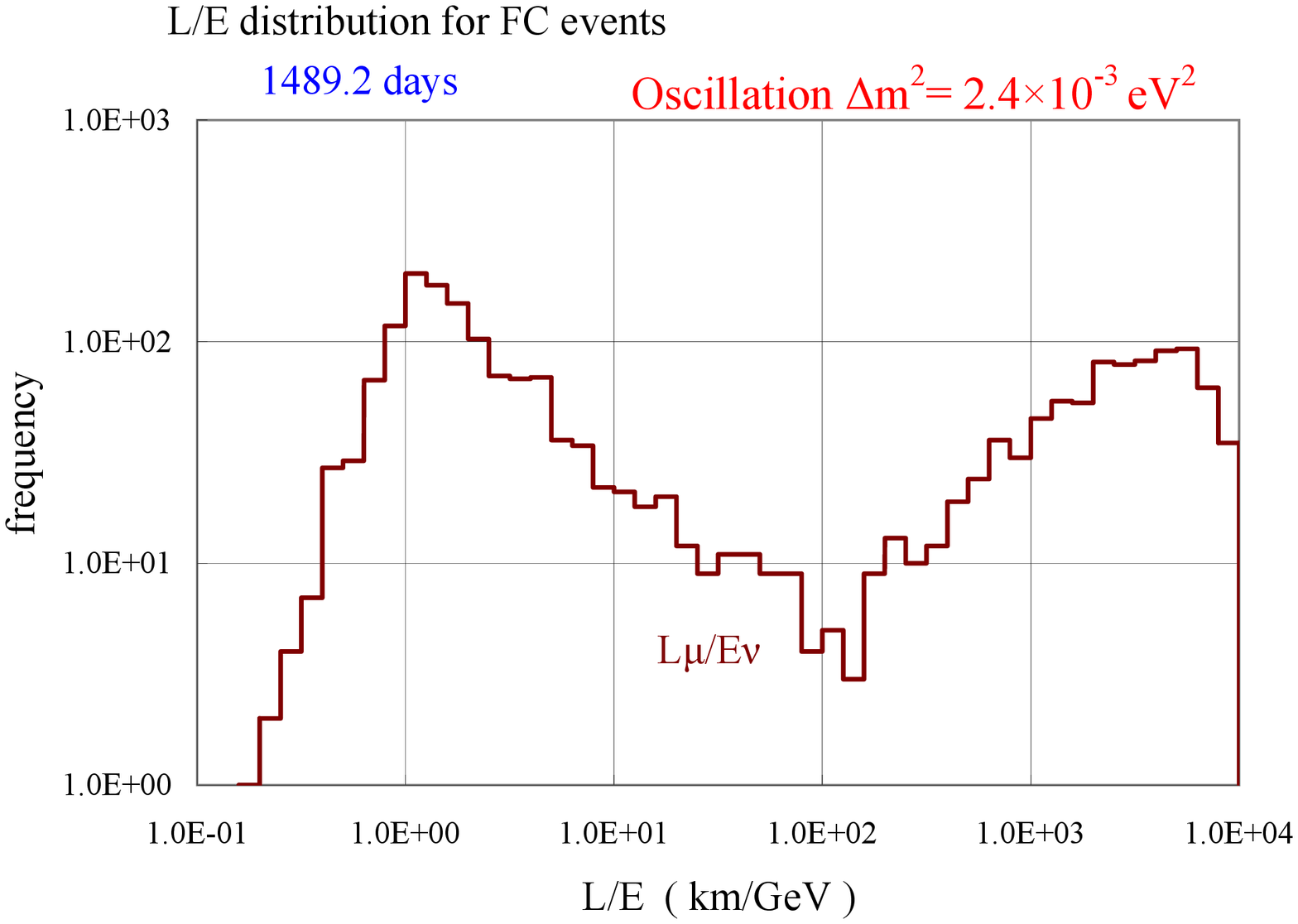}
}}
\vspace{-1cm}
\caption{$L_{\mu}/E_{\nu}$ distribution with the oscillation
for 1489.2 live days.}
\label{fig:21}
\vspace{-1cm}
\rotatebox{90}{%
\resizebox{0.5\textwidth}{!}{%
  \includegraphics{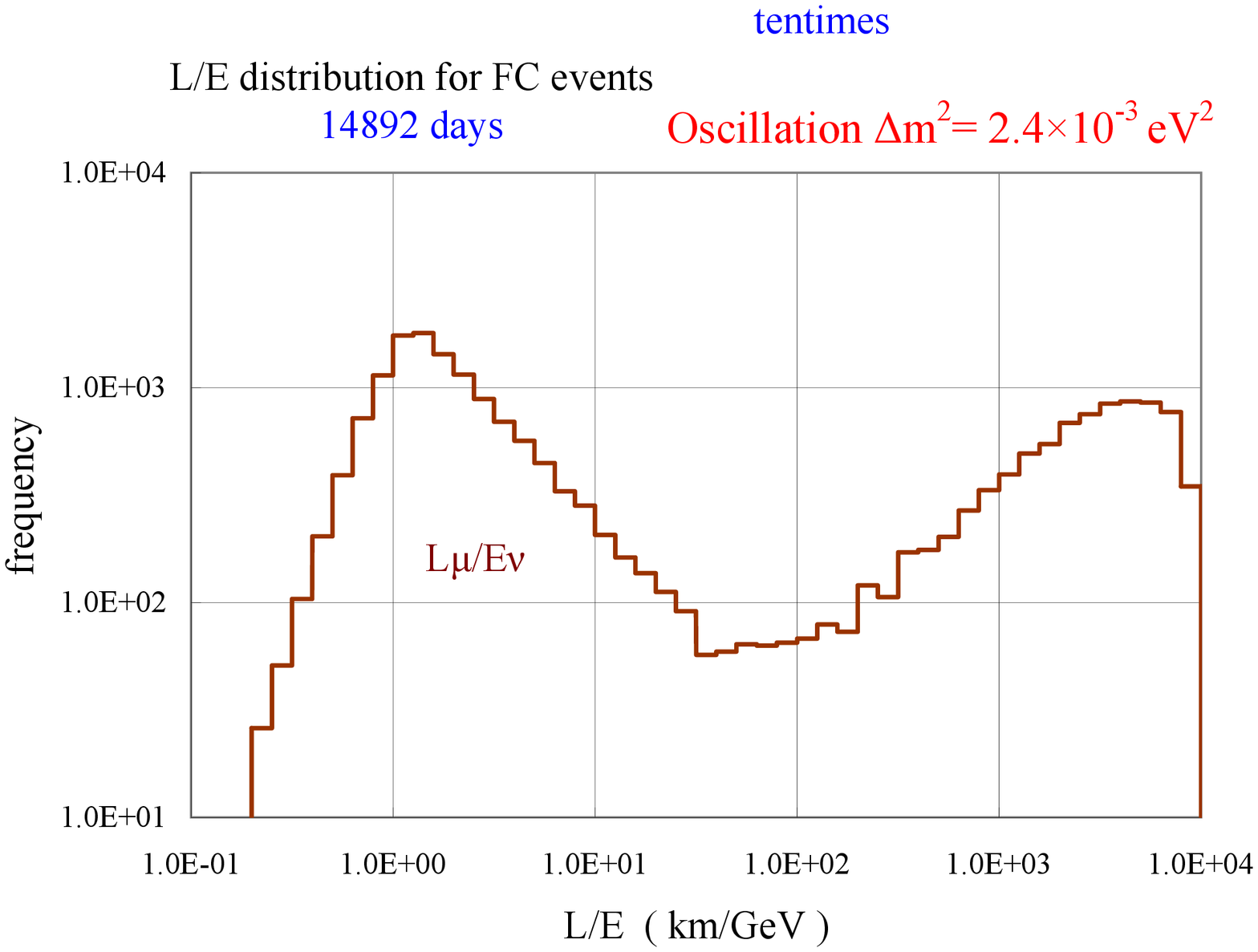}
}}
\vspace{-1cm}
\caption{$L_{\mu}/E_{\nu}$ distribution with the oscillation
for 14892 live days.}
\label{fig:22}
\end{center}
\end{figure}
\begin{figure}
\begin{center}
\vspace{-1cm}
\rotatebox{90}{%
\resizebox{0.5\textwidth}{!}{%
  \includegraphics{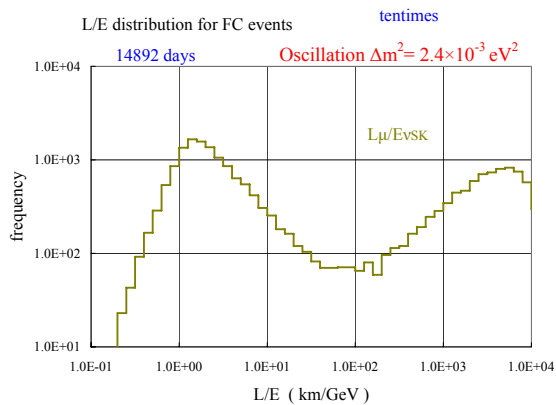}
}}
\vspace{-1cm}
\caption{The $L_{\mu}/E_{\nu,SK}$ distribution with oscillation 
for 14892 day.}
\label{fig:23}
\end{center}
\end{figure}
\begin{figure}
\begin{center}
\vspace{-1cm}
\rotatebox{90}{%
\resizebox{0.5\textwidth}{!}{%
  \includegraphics{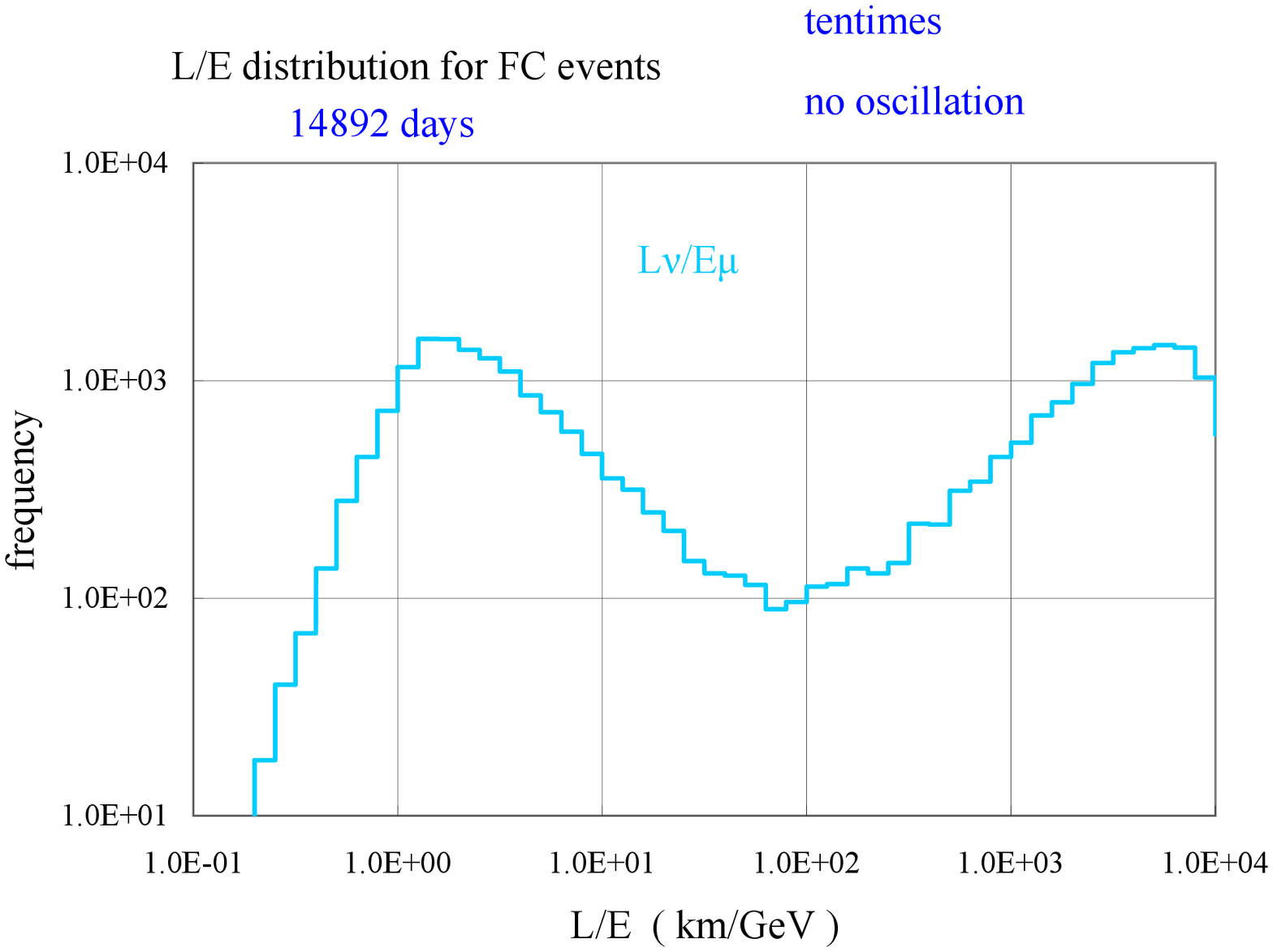}
}}
\vspace{-1cm}
\caption{The $L_{\nu}/E_{\mu}$ distribution without
oscillation for 14892 days.}
\label{fig:241}
\vspace{-0.5cm}
\rotatebox{90}{%
\resizebox{0.5\textwidth}{!}{%
  \includegraphics{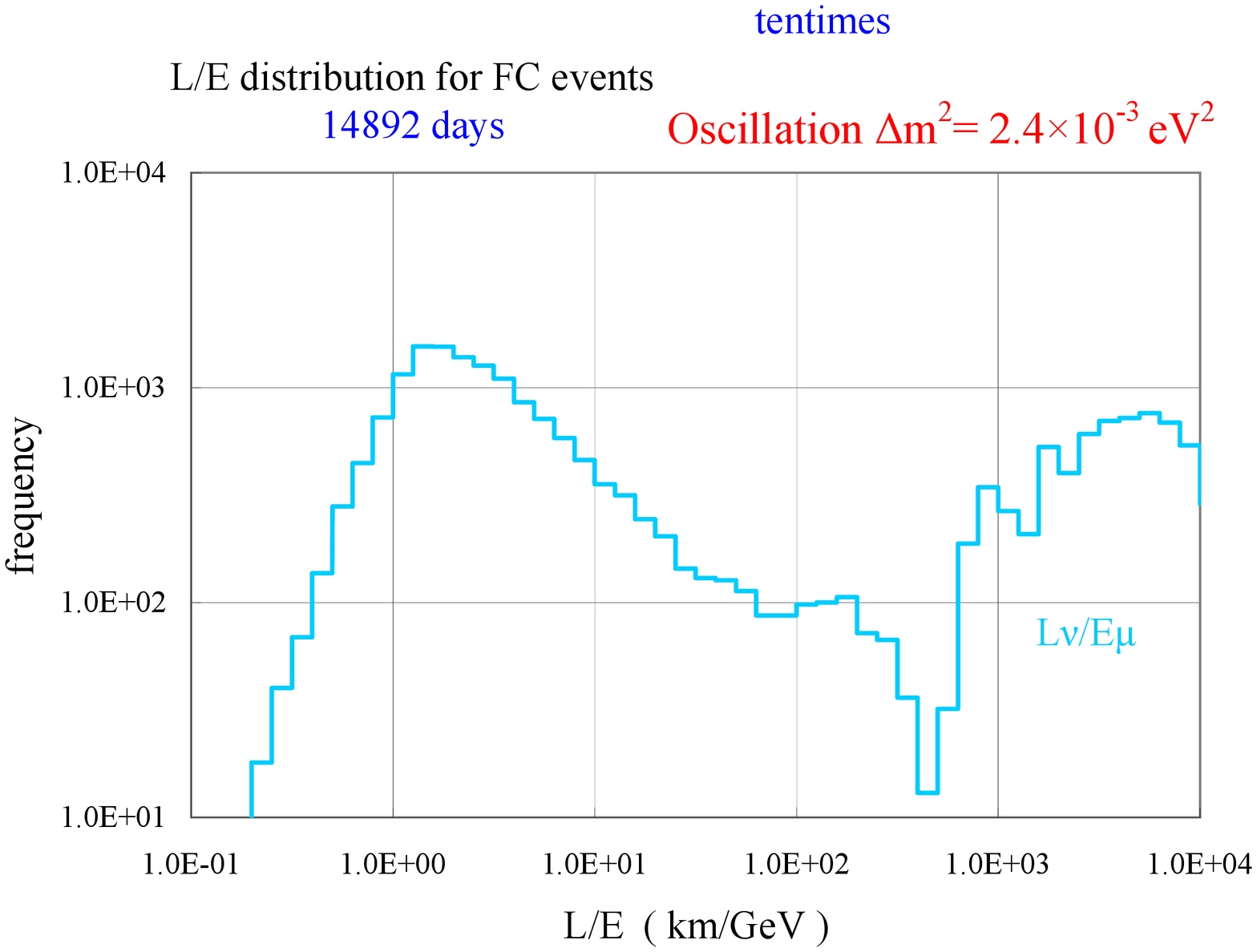}
}}
\vspace{-1cm}
\caption{The $L_{\nu}/E_{\mu}$ distribution with the
oscillation for 14892 days.}
\label{fig:24}
\end{center}
\end{figure}
In order to confirm the disappearance of the psuedo maximum 
oscillations, in Figures~7 and 8, we give
the survival probability of a given flavor 
for $L_{\mu}/E_{\mu}$ distribution, namely,\\
$(L_{\mu}/E_{\mu})_{osc}/(L_{\mu}/E_{\mu})_{null}$,
 for 1489.2 live days and 14892 live days, respectively.
 Comparing Figure~7 with Figure~8, pseudo dips in Figure~7 
disappear in Figure~8.
Thus the histogram becomes a rather decreasing function of 
$L_{\mu}/E_{\mu}$ in Figure~8. 
If we further make statistics higher, the survival probability for
$L_{\mu}/E_{\mu}$ distribution should be a monotonously decreasing 
function of $L_{\mu}/E_{\mu}$, whithout showing any 
characteristics of the maximum oscillation,  
which is contrast to Figures~8, 9 and 10
in the preceeding paper\cite{part2}.
 In conclusion, we should say that we can not find any maximum 
oscillation for the neutrino oscillation in the 
$L_{\mu}/E_{\mu}$ distribution.  
  

\subsection{$L_{\mu}/E_{\nu}$ Distribution}
Now, we examine the $L_{\mu}/E_{\nu}$ distribution which 
the Super-Kamiokande Collaboration treat 
in the thier 
paper, expecting the evidence for the oscillatory signatuture in 
atmospheric neutrino oscillations.
\subsubsection{For null oscillation}
In Figures~9 and 10, we give the $L_{\mu}/E_{\nu}$ distribution without 
oscillation for 1489.2 live days and 14892 live days, respectively.
Comparing Figure~9 with Figure~10, the larger statistics makes the
 distribution more smooth.
 Also, there is sinusoidal-like dip which have no relation with 
neutrino oscillation.
\subsubsection{For the oscillation}
In Figures~11 and 12, we give the $L_{\mu}/E_{\nu}$ distribution with 
oscillation for 1489.2 live days and 14892 live days, respectively. 
In Figure~11, we may find something like dip which corresponds to the 
first maximum oscillation near $\sim$200 (km/GeV).
However, such the dip disappears, by making the statistics larger as 
shown in Figure~12. 
Instead, Figure~12 gives the histogram with a little unnatural shape 
 in spite of 
larger statistics. This may come from the complicated correlation 
between $L_{\mu}$ and $E_{\nu}$, the details of which 
are shown partially in Eq.(2), Eq.(3) and Figure~5
in the preceeding paper\cite{part2}. 

\subsubsection{$L_{\mu}/E_{\nu,SK}$Didtribution for the oscillation}
 Instead of $E_{\nu}$ which is correctly sampled from the corresponding 
probability functions, 
let us utilize $E_{\nu,SK}$ which is obtained from the "approximate" 
formula (Eq.(4) in the preceeding paper\cite{part2}).
 We express $E_{\nu}$ described in Eq.(4) in the preceeding paper\cite{part2} utilized by the
Super-Kamiokande Collaboration as $E_{\nu,SK}$ 
to discriminate our $E_{\nu}$ obtained in stochastic manner 
correctly.
In Figure~13, we give $L_{\mu}/E_{\nu,SK}$ distribution for 14892 
live days and 14892 live days, ten times as much as the 
Super-Kamiokande Experiment actual live days.  
. 
If we compare Figure~13 with Figure~12, we understand that there are 
no significant difference between them. 
This fact tells us that the "aproximate" formula for $E_{\nu}$ by 
the Super-Kamiokande Collaboration, 
which is not suitable 
for the treatment of the stochastic quantities, does not produce so 
significant error actually,
which is understandable from Figure~5
in the preceeding paper\cite{part2}.
 Also, we can conclude that we do not 
find any dip corresponding to any maximum oscillation from 
$L_{\mu}/E_{\nu}$ or $L_{\mu}/E_{\nu,SK}$ distributions.
The reason why the Figures 10 and 13 can not show any dip structure, 
which is shown in Figures from 8 to 10 
in the preceeding paper\cite{part2} clearly, 
 comes from the situation that the role of $L_{\nu}$
is much more crucial than that of $E_{\nu}$ in the $L/E$ analysis.
Namely, $L_{\nu}$ cannot be replaced by $L_{\mu}$ at all.
 Also, see the discussion in the following subsection 2.3.    
    
\begin{figure}
\begin{center}
\vspace{-1cm}
\rotatebox{90}{%
\resizebox{0.5\textwidth}{!}{%
  \includegraphics{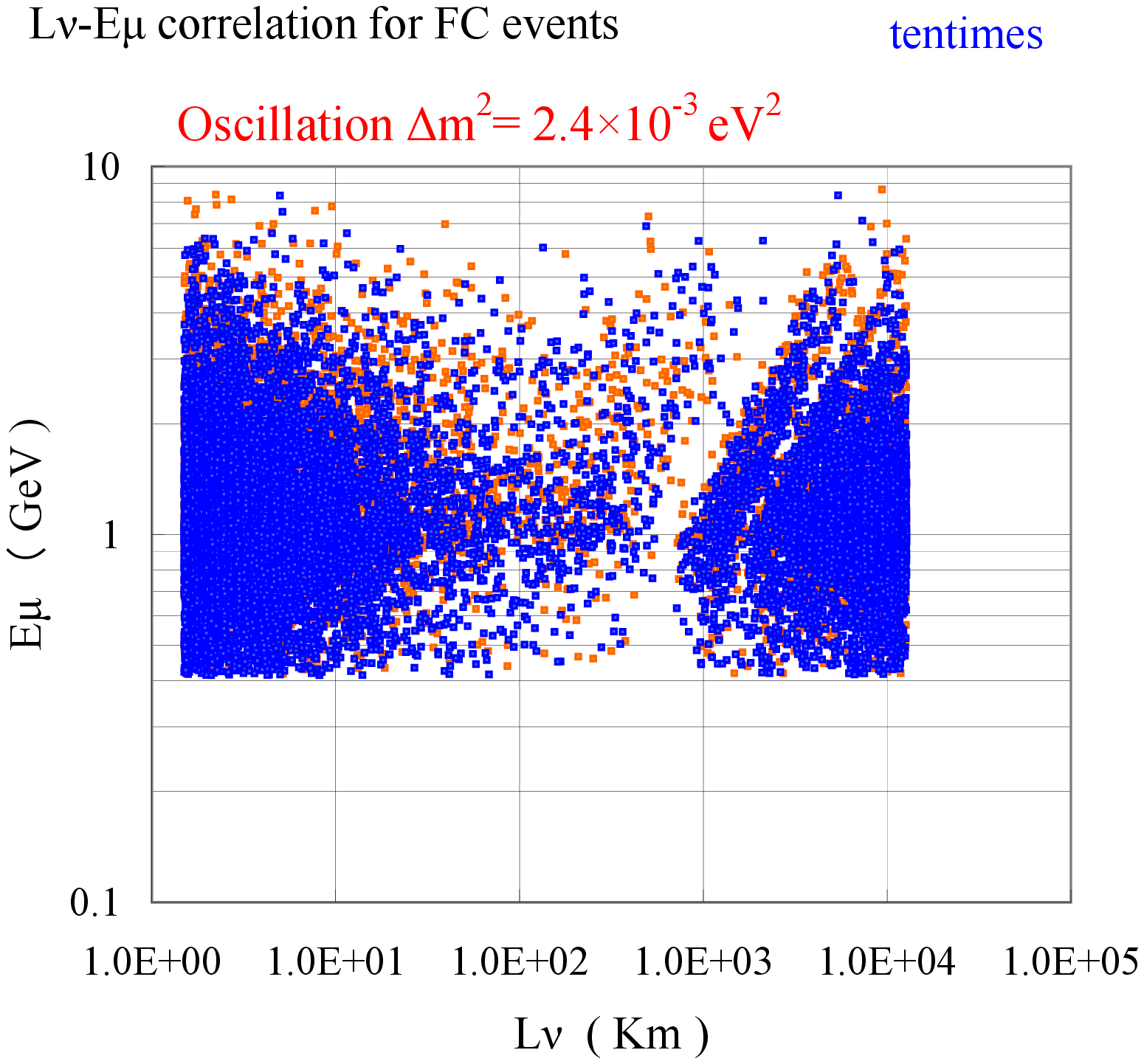}
}}
\vspace{-1cm}
\caption{The correlation diagram between $L_{\nu}$ and $E_{\mu}$
 with the oscillation for 14892 days.}
\label{fig:25}
\vspace{-1cm}
\rotatebox{90}{%
\resizebox{0.5\textwidth}{!}{%
  \includegraphics{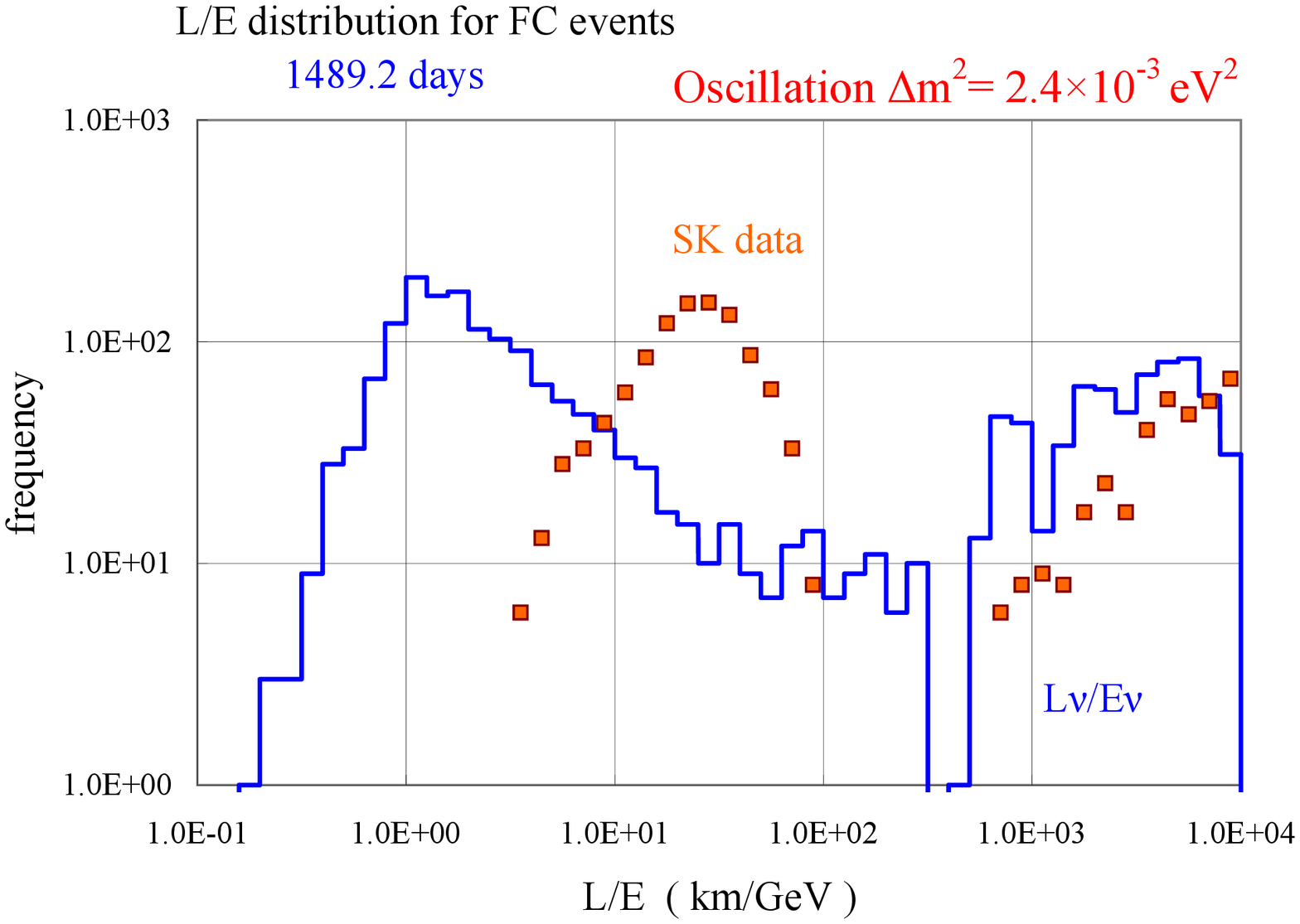}
}}
\vspace{-1cm}
\caption{The comparison of $L/E$ distribution for single-ring muon 
events due to QEL among {\it Fully Contained Events} with the 
corresponding one by the Super-Kamiokande
Experiment.}
\label{fig:26}
\end{center}
\end{figure}

\subsection{$L_{\nu}/E_{\mu}$ Distribution}
\subsubsection{For null oscillation}
 In Figure~14, we give $L_{\nu}/E_{\mu}$ distribution without 
oscillation for 14892 days, ten times as much as actual live days
of the Super-Kamiokande Experiment
 to consider statistical fluctuation effect as precisely as possible. 
It is clear from the figure that there is not any dip 
corresponding to 
the maximum oscillation which is expected to appear in the presence of 
the neutrino oscillation. 
\subsubsection{For the oscillation}
In Figure~15, we give the corresponding  
distribution with the oscillation. In Figure~16, we give the correlation 
diagram between $L_{\nu}$ and $E_{\mu}$which correspond to Figure~15. 
On the contrary to Figure~14, there are surely dips in Figure~15,
 and furthermore we can discriminate the strip pattern in Figure~16, 
similarly as in the Figure~12 in the preceeding paper\cite{part2}.

Therefore, we suppose from Figures~15 and 16 that we may observe 
some quantities which is directly related to the maximum 
oscillations in the $L_{\nu}/E_{\nu}$ distribution.
 However, it seems to be difficult to extract a pair of concrete 
values of $L_{\nu}$ and $E_{\nu}$ through the analysis of
$L_{\nu}/E_{\mu}$ distribution.
 Comparing Figure 4 with Figure 5
in the preceeding paper\cite{part2}, it is clear that
$L_{\nu}$ can not be approximated by $L_{\mu}$ at all,
 while $E_{\nu}$ can be approximated by  $E_{\mu}$ within some 
allowance (see Figure~5 in the preceeding paper\cite{part2}
).
Thus, the $L_{\nu}/E_{\mu}$ distribution can show some 
similar structure to $L_{\nu}/E_{\nu}$ distribution.
 This fact shows that the role of $L_{\nu}$ is essentially important  
compared with $E_{\nu}$ in the $L/E$ analysis. 
However, it should be noticed again that we can not observe 
the $L_{\nu}/E_{\mu}$ distribution physically even if the 
dips surely exist in this distribution, because $L_{\nu}$ is 
the physically unobservable quantity.   

\section{Comparison of distribution from the 
Super-Kamiokande Experiment with our results}
As the Super-Kamiokande Collaboration 
think that they can approximate $L_{\mu}$ nearly equal to 
$L_{\nu}$ and $E_{\nu}$ is well approximated by Eq.(4),
 their experimental data should be compared with our 
$L_{\nu}/E_{\nu}$ distribution.\\ 
In Figure~17, we compare our numerical experimental data 
for {\it Fully Contained Events} due to QEL with the 
corresponding one by the Super-Kamiokande
Experiment (read from Figure 8.22 \cite{Ishitsuka}).  
In the light of the correct distribution, uncertainties in the
distribution from the Super-Kamiokande Experiment
consist of uncertainty in $L_{\mu}$ (see Figure~4 and 
Eq.(3) in the preceeding paper\cite{part2}) 
and in the transformation of $E_{\nu}$ from $E_{\mu}$
(see Figure~5 in the preceeding paper\cite{part2}). There are big 
differences between 
our distribution and the corresponding one from 
the Super-Kamiokande Experiment.
 The first is the difference in the shape of the distribution and 
the second is in their dip structure.
 It seems to be curious that there exists a rather wider dip 
from 100 to 630~km/GeV for the first maximum oscillation in the 
distribution from the Super-Kamiokande Experiment,
 which is against the sense of maximum oscillation,
while we give a sharp dip for the first maximum oscillation 
around 520~km/GeV predicted by the neutrino oscillation parameters
from the Super-Kamiokande Collabolation.  
 In order to clarify the reason for the remarkable difference 
between ours and that of the Super-Kamiokande Experiment,
 it is required that 
the Super-Kamiokande Collaboration disclose their correlation 
diagram between $L_{\nu}$ and $E_{\nu}$ as shown in Figure 12
in the preceeding paper\cite{part2}.

\section{Conclusion}
  The Super-Kamiokande
Collabolation trys to get the evidence for an oscillatory signature in 
atmospheric neutrino oscillations by detecting the maximum oscillations 
(the first maximum oscillation). 
Then, they approximate $L_{\nu}$ by $L_{\mu}$ and estimate $E_{\nu}$
from $E_{\mu}$ in their $L/E$ analysis. However, we show that the 
approximation of $L_{\nu}$ by $L_{\mu}$ 
doest not hold at all (Figures~3 and 4 in the present paper) and the
 estimation method by the Super-Kamiokande Collabolation
in energy is theoretically unsuitable (Figure~5 in the 
preceeding paper\cite{part2}). 
Then, it is clarified that the role of $L_{\nu}$ is  
more decisively cruisial than that of $E_{\nu}$ in the $L/E$ analysis. 
As a result of it, one can not replace 
$L_{\nu}/E_{\nu}$  by $L_{\mu}/E_{\nu}$.

In the $L/E$ analysis, we examine all possible combinations of $L/E$,
 namely, 
$L_{\nu}/E_{\nu}$\cite{part2}, $L_{\nu}/E_{\mu}$, $L_{\mu}/E_{\nu}$
and $L_{\mu}/E_{\mu}$ in the present paper.
Among all possible $L/E$ analysis, we find only 
the $L_{\nu}/E_{\nu}$ distribution can give the maximum oscillations 
from the survival probability of a given flavor (Eq.~1)), as it must be. 
However, the $L_{\nu}/E_{\nu}$ distribution can not be 
physically observed. Even if we put aside the unsuitable estimation 
of $E_{\nu}$ from $E_{\mu}$ by the Super-Kamiokande
Collabolation(Eq.~4 in the preceeding paper\cite{part2}),
 it is concluded from our 
analysis by the numerical computer experiment
that $L_{\mu}/E_{\nu}$ distribution by the Super-Kamiokande
Collabolation can not 
obtain the maximum oscillation from the survival 
probability of a given flavor. 
 From the experimental point of view, physically measurable 
quantities are $L_{\mu}$ and $E_{\mu}$. Therefore, it is desirable that 
the Super-Kamiokande Collaboration 
carry out the $L_{\mu}/E_{\mu}$ analysis from which they 
examine whether they can really observe the maximum oscillation 
for neutrino oscillation or not.
In this case, we are free from the uncertainty which is produced by 
the estimation of $E_{\nu}$ from $E_{\mu}$.
 However, even if the Super-Kamiokande
Collabolation utilizes $E_{\mu}$ instead of $E_{\nu}$,
 we can not observe the maximum oscillation in the $L_{\mu}/E_{\mu}$
 analysis, which are shown in Figure~8.\\ 
 
Furthermore, it should be emphasized that confirmation 
of the existence of the maximum oscillations can be carried out 
by the analysis on the ratio of
$(L/E)_{osc}/(L/E)_{null}$, but not by that of the $(L/E)_{osc}$ only. 
For the purpose, we should say the 
numerical computer experiment is an indispensable mean.   
In conclusion, we would say that we can not observe any maximum
oscillations with the Super-Kamiokande Experiment
 $L/E$ analysis
against the original claim by the Super-Kamiokande
Collabolation.

\section*{References}



\begin{thebibliography}{}
%
%
  \bibitem{part2} Konishi,E.,Minorikawa,Y.,Galkin,V.I.,Ishiwata,M.,
Nakamura,I.,Kato,M. and Misaki,A arXiv:hep-ex/0808.3313v1
  \bibitem{Renton} Renton, P., {\it Electro-weak Interaction}, Cambridge University Press (1990). See p. 405.
  \bibitem{Ashie2} Ashie,Y {\it et al.}, Phys.Rev.{\bf D171}(2005)112005

  \bibitem{Ashie1} Ashie,Y {\it et al.}, Phys.Rev.Lett.{\bf93}(2004)101801-1

  \bibitem{Ishitsuka} Ishitsuka, M., PhD thesis, University of Tokyo (2004). See p. 138.



\end{thebibliography}
\end{document}